\documentstyle[12pt]{article}
\def\double{\baselineskip 24pt \lineskip 10pt}
\begin{document}
\begin{titlepage}
\begin{flushright}
BA-95-19; astro-ph/9507006
\end{flushright}
\begin{center}
\Large
{\bf Cold Plus Hot Dark Matter \\
Cosmology in the Light of \\
Solar and Atmospheric\\
Neutrino Oscillations\footnote{Work supported in part
by grants from the DOE and NASA.} }\\
\vspace{.6cm}
\normalsize
\large
K.S. Babu\footnote{ Address starting September 1995: School of Natural
Sciences, Institute for Advanced Study, Olden Lane, Princeton, NJ 08540}, R.K.
Schaefer, and Q. Shafi \\
\normalsize
\vspace{.2 cm}
{\em Bartol Research Institute \\
University of Delaware \\
Newark, Delaware 19716 }
\vspace{.3 cm}
\end{center}
\begin{abstract}
\noindent
We explore the implications of possible neutrino oscillations, as indicated
by the solar and atmospheric neutrino experiments, for the cold plus hot
dark matter scenario of large scale structure formation.  We find that
there are essentially three distinct schemes that can accommodate the
oscillation data and which also allow for dark matter neutrinos. These
include (i) three nearly
degenerate (in mass) neutrinos, (ii) non-degenerate masses with $\nu_\tau$ in
the eV range, and (iii) nearly degenerate $\nu_\mu-\nu_\tau$ pair (in the eV
range), with the additional possibility that the electron neutrino is
cosmologically significant.  The last two schemes invoke a
`sterile' neutrino which is light
($\mathrel{\raise.3ex\hbox{$<$\kern-.75em\lower1ex\hbox{$\sim$}}}$ eV).
We discuss the implications of these schemes for $\bar{\nu}_\mu -
\bar{\nu}_e$ and $\nu_\mu - \nu_\tau$ oscillation, and find that
scheme (ii) in particular, predicts them to be in the observable range.   As
far as structure formation is concerned, we compare the one neutrino
flavor case with a variety of other possibilities, including two and
three degenerate neutrino flavors.  We show, both analytically and
numerically, the effects of these neutrino mass scenarios
on the amplitude of cosmological density fluctuations.  With a Hubble
constant of 50 km s$^{-1}$ Mpc$^{-1}$, a spectral
index of unity, and $\Omega_{baryon} = 0.05$, the two and three flavor
scenarios fit the observational data marginally better than the single
flavor scheme.  However, taking account of the uncertainties in these
parameters, we show that it is premature to pick a clear winner.

\end{abstract}
\vspace{-.3cm}
\begin{center}
{\small PACS numbers: \hspace{0.5cm} 98.65-r, 14.60-st, 95.35+d}\\
\end{center}
\end{titlepage}
\double

\section{Introduction}

The idea that the dark matter in the universe may contain both a `cold' as
well as a `hot' component \cite{ss84} first arose as a serious possibility
\cite{gutcphdm} within the
framework of grand unified theories  (which also inspired inflation, cosmic
strings, baryogenesis, etc.). In the original models the cold component was
the axion, although nowadays the lightest supersymmetric particle (LSP) often
plays that role.  Lightly massive neutrinos are the hot
component, and it was usually assumed that the neutrinos would display the
hierarchical
mass spectrum characteristic of the quarks and charged leptons, with a single
neutrino ({\it e.g.}, $\nu_\tau$) dominating the mass density of the hot
component.

 However, it was noted from the beginning \cite{ss84} that because of
their special nature, nearly degenerate massive neutrinos is a logical,
albeit not necessarily the simplest, possibility.
The cosmological implications of neutrinos that are closely
degenerate in mass has received much recent
attention \cite{pogosyan95,primack95,also}, inspired to some extent by the
remarkable series of neutrino oscillation
experiments (solar, atmospheric, and more recently accelerator) which
suggest that two or more
neutrino flavors may contribute to the hot dark matter in the universe.

Before discussing the impact of neutrino oscillation experiments on the
cold plus hot dark matter (C$+$HDM) scenario, let us briefly recall why the
latter has attracted so much recent attention.  From the mid to the
late 1980s evidence was mounting that the so-called ``standard" cold
dark matter (CDM) scenario had trouble providing a consistent explanation of
small ($\sim$ galactic) as well as large scale structure.  On the contrary, a
C$+$HDM model (with $\Omega_{\nu} \sim 0.15 - 0.35$) provided a far more
consistent fit to the data \cite{vdns}.  In order to provide additional
tests, the quadrupole anisotropy of the microwave background expected in
C$+$HDM models was estimated in 1989 \cite{sss89,holtzman89} and compared
with the CDM prediction.  Normalized to the
`small' scale data, it was found that the C$+$HDM prediction exceeded the CDM
value by a factor of about two, and this was dramatically verified when the
COBE team \cite{smoot92} made its announcement in 1992 \cite{shafi}.

In the last two years, particularly after extensive numerical simulations of
galaxy and cluster formation \cite{nbody} and more detailed analytical work
\cite{others,pogosyan93}, additional support for the C$+$HDM model has
emerged.  It appears that this
model provides the simplest (consistent) realization of an inflationary
scenario
for  large scale structure formation.
%
%
It has recently been pointed out \cite{pogosyan95,primack95}, (and our results
are in agreement) that with two or more neutrino flavors contributing to
the hot dark matter, somewhat better fits to the present data may be
possible.  For fixed $\Omega _{\nu}$ there is also some dependence both on the
spectral index as well as the Hubble constant.  It must be admitted, though,
that within the framework of grand unified theories, it may not be easy to
realize scenarios with two or three nearly degenerate  (in mass) neutrinos.
Furthermore,  in some of the scenarios we consider, one needs to invoke a
fourth
(sterile)  neutrino which is nearly degenerate in mass with the electron
neutrino.  Precisely why this singlet state can be so light is an important
question for future research.

The layout of the paper is as follows.  In section 2.1 we list three sets of
observations related to solar, atmospheric, and dark matter neutrinos which
provide hints for non-zero neutrino masses.  For completeness, we also
summarize the results of the LSND experiment at Los Alamos, although we will
not include them among the constraints to be satisfied.  We then proceed to a
discussion of three distinct schemes which are consistent with the observed
solar and atmospheric $\nu$ oscillations, and which also allow for neutrino
dark matter.  We discuss the implications of these schemes for a variety of
experiments, particularly the $\bar{\nu}_\mu-\bar{\nu}_e$ and
$\nu_\mu - \nu_\tau$ oscillation experiments at accelerators.
In section 3 we address the main issues raised by
neutrino oscillations for cosmological structure formation within the C$+$HDM
framework.  We review the physics of the differences between having one and
more than one massive neutrino flavor, giving some analytic formula to estimate
the scale (section 3.1) and amplitude (section 3.2) of the effects on the
growth of primordial density fluctuations.  In 3.3 we give results
of some more detailed numerical computations of the cosmological models based
on the solutions to the oscillation data, including some models, in 3.4,
motivated by the LSND results.
Finally, in section 3.5 we discuss the dependence on the cosmological
parameters (Hubble constant, spectral index $n$, and baryonic density
fraction) of conclusions drawn from comparing specific models to large scale
structure data.

\section{ Neutrino Masses and Mixings}

\subsection{Observational Hints}

One of the greatest challenges in particle physics today pertains to the issue
of the neutrino masses.  There are indications from a variety of experiments
that one (or more) of the known neutrinos may possess non--zero mass(es).  In
this section we briefly summarize the relevant observations based on solar and
atmospheric neutrinos, as well as present arguments for neutrino dark matter.
[For completeness, we also summarize the recent findings of the LSND
experiment at Los Alamos.]  In section 2.2, we discuss  scenarios for neutrino
masses and mixings which can simultaneously accommodate these
observations.  We find that there are essentially three viable schemes, and
remarkably enough, all three will be tested in ongoing and planned neutrino
oscillation experiments (particularly Los Alamos LSND,
Rutherford KARMEN, the CERN
CHORUS/NOMAD and Fermilab E803 experiments).

\begin{enumerate}
\item {\it Solar neutrino puzzle:}  The apparent deficit in the flux
of solar $\nu_e$'s \cite{Bahcall}
that has persisted in the Chlorine experiment for
over two decades \cite{Cl} has been confirmed
in the last several years by three independent
experiments: the Kamiokande water Cerenkov detector \cite{KII},
and the SAGE \cite{SAGE} and
GALLEX \cite{GALLEX} radiochemical experiments.  These four
observations, which probe different energy regimes in the
solar neutrino spectrum,
can all be simultaneously explained in terms of two flavor neutrino
oscillations.  The matter enhanced MSW \cite{MSW} oscillation for a two flavor
($\nu_e-\nu_x$) system admits two branches \cite{Langacker}: (i) The small
angle
non--adiabatic solution, which requires sin$^22\theta = (3.5 \times 10^{-3}
{}~{\rm to}~ 1.5 \times
10^{-2})$ and $\Delta m^2 \equiv (m_2^2-m_1^2) = (3.4 \times 10^{-6}
{}~{\rm to}~ 1.2 \times 10^{-5})~eV^2$.
(ii) The large angle MSW solution requiring sin$^22\theta =
(0.6-0.9)$ and $\Delta m^2 = (7 \times 10^{-6}~{\rm to}~5 \times 10^{-5})
{}~eV^2$.  Here $\nu_x$ could be
$\nu_\mu, \nu_\tau$ or a sterile neutrino $\nu_s$ for the small angle
MSW solution, while $\nu_x= (\nu_\mu~ {\rm or}~ \nu_\tau)$
in the large angle MSW solution.
There is also the $(\nu_e-\nu_x)$ vacuum oscillation solution \cite{vacuum}
for $\nu_x = (\nu_\mu ~{\rm or}~\nu_\tau)$, with sin$^22\theta =
(0.6-1.0)$ and $\Delta m^2 \sim 10^{-10}~eV^2$.

\item {\it Atmospheric neutrino anomaly:}  The ratio of muon to
electron neutrinos from atmospheric cascades measured by
the Kamiokande \cite{Kam} and the IMB \cite{IMB}
experiments appear to suggest a deficit
of about a factor of 2 when compared to Monte Carlo
simulations \cite{Gaisser}.  For sub--GeV neutrinos (i.e., neutrinos with
energy
less than $\sim $ 1 GeV), Kamiokande finds for this ratio of ratios
(which is expected to be 1 in the absence of neutrino oscillations)
$R = 0.60^{+0.06}_{-0.05} \pm 0.05$ which is in good agreement with
the IMB value of $R = 0.54 \pm 0.05 \pm 0.12$.  Recently Kamiokande has
also reported results for the multi--GeV neutrinos \cite{Fukuda}
where the ratio is
$R = 0.57^{+0.08}_{-0.07}\pm 0.07$, in agreement with the sub-GeV data.
This apparent deficit of $\nu_\mu$'s or the excess of $\nu_e$'s can be
attributed to either $(\nu_\mu-\nu_\tau)$ oscillations with sin$^22\theta
= (0.65-1.0)$ and $\Delta m^2 =(5 \times 10^{-3}~{\rm to}~3 \times
10^{-2})~eV^2$, or to
$(\nu_e-\nu_\mu)$ oscillations with sin$^22\theta = (0.55-1.0)$ and
$\Delta m^2 = (7 \times 10^{-3}~{\rm to}~7 \times 10^{-2})~eV^2$.
The resolution of the anomaly in terms of $(\nu_\mu-\nu_s)$
oscillations, where $\nu_s$ is a sterile neutrino, will run into
difficulty with primordial nucleosynthesis calculations which require
that \cite{nuc}
$\Delta m^2 {\rm sin}^22\theta \le 1.6 \times 10^{-6}~eV^2$.  (Note
that the solar $(\nu_e-\nu_s)$ oscillation parameters satisfy this
constraint.)  Among the other atmospheric neutrino experiments,
only Soudan II \cite{Soudan} sees an anomaly with a preliminary value of
$R = 0.69 \pm 0.19 \pm 0.09$.  The Frejus \cite{Frejus} results are
marginally in conflict with the combined Kamiokande and
IMB data.  Considering the low statistics of Frejus experiment
compared with the high statistical significance of Kamiokande and IMB data,
we shall ignore this marginal discrepancy in our theoretical discussions.

\item {\it Hot component of dark matter:}  As already noted in the
introduction, and further explained in the next section, a combined fit
to the COBE data as well as the data on
distribution of galaxies on the
large and small angular scales is difficult to achieve within the cold dark
matter scenario.  The simplest consistent scenario requires a
significant hot component (15-30)\% in the dark matter
\cite{pogosyan93,ss94,dss}
leading to the
C$+$HDM scheme, with neutrinos being the natural candidates for the hot
component.  The mass of the neutrino comprising the hot dark matter
should be in the few eV range.

\item {\it $\overline{\nu}_\mu-\overline{\nu}_e$ oscillation at
Accelerators:}  Recently the LSND experiment at the LAMPF
facility in Los Alamos has reported positive evidence for
$\overline{\nu}_\mu-\overline{\nu}_e$ oscillations \cite{LSND}.
If these results survive further scrutiny and there is independent
verification (e.g, at the KARMEN experiment \cite{KARMEN}),
it certainly will have a strong
impact on particle and nuclear physics as well as cosmology and
astrophysics.   The initial LSND data, if interpreted in terms of
two--neutrino oscillations suggest a mixing probability of
$P_{\overline{\nu}_\mu\rightarrow \overline{\nu}_e}=
(0.34^{+0.20}_{-0.18}\pm 0.07)\%$, when oscillation constraints from
KARMEN and BNL-E776 \cite{BNL} accelerator searches as well as reactor neutrino
constraints from Bugey facility \cite{Bugey} are folded in.
In the discussion that follows, we shall take into considerations the
sensitivity of the LSND and other accelerator and reactor
experiments in searching for
$\overline{\nu}_\mu-\overline{\nu}_e$ oscillations, but we shall not
demand that the theoretical scenarios discussed correctly reproduce the
precise numerical values reported by the LSND collaboration.  We feel
that this position is justified at this time, especially in view of the
fact that a different analysis of the LSND data has yielded considerably
weaker limits on the mixing parameters \cite{Hill}.
We find it remarkable however, that independent of the LSND data,
two of the three scenarios which can
accommodate points (1)--(3) above, imply an observable
signal at accelerator $(\overline{\nu}_\mu-\overline{\nu}_e)$ oscillation
experiments.

\end{enumerate}

\noindent {\bf 2.2 Theoretical Schemes:}

Let us focus on the solar, atmospheric and
hot dark matter neutrinos, leaving out the LSND results which will not play a
role in determining the viable scenarios.
It is clear that the mass--splittings required for explaining these
three observations do not overlap ($\Delta m^2 \sim 10^{-5}~eV^2$ for
solar neutrinos, $\Delta m^2 \sim 10^{-2}~eV^2$ for atmospheric
neutrinos and $m_\nu \sim$ (1 to few) eV for hot dark matter).
Assuming that there are no additional light neutrinos,
one concludes that the three neutrinos $(\nu_e,\nu_\mu,\nu_\tau)$ should be
nearly degenerate in mass.  (Note that we often identify a mass eigenstate
through its dominant flavor.  The respective masses are $m_1,\ m_2,$ and
$m_3$.)  This leads us to our first scenario.

\noindent \underline{{\it Scenario (i):
Three nearly degenerate neutrinos:}}
Assume that
$\nu_e,\nu_\mu$ and $\nu_\tau$ have a nearly
common mass of about (1 to 3) eV.  Their masses are split by small amounts,
such
that $\Delta m_{12}^2 \equiv (m_2^2-m_1^2) \sim 10^{-5} eV^2$ and
$|\Delta m_{23}^2| \equiv |(m_3^2 - m_2^2)| \sim 10^{-2}~eV^2$.
Phenomenological neutrino mass matrices that are consistent with these
assumptions are easily constructed \cite{caldmohap}.
This scheme will account for the
solar neutrino data via $(\nu_e-\nu_\mu)$ MSW oscillations and the
atmospheric neutrino anomaly in terms of $(\nu_\mu-\nu_\tau)$
oscillations.  The required mixing angles are free parameters and can be
adjusted to the suggested values.  The $(\nu_e-\nu_\tau)$ mixing angle
can be arbitrarily small, but it can also be as large as about 0.05.
According to this scenario, neutrinoless double beta decay ($\beta
\beta_{0\nu}$)  is at (or even above?) the
present experimental limit (modulo nuclear matrix element uncertainties), and
so that it can be ruled out if $\beta \beta_{0\nu}$ is not observed in
the near future.

Also according to this scheme the $\nu_\mu-\nu_\tau$ oscillation experiments
at accelerators
(CHORUS/NOMAD, E803 \cite{Chorus,Nomad}),
as well as $\nu_e-\nu_\mu$ oscillation
searches at reactors, should fail to find anything interesting.  In both cases
the relevant $\Delta m^2
\mathrel{\raise.3ex\hbox{$<$\kern-.75em\lower1ex\hbox{$\sim$}}} 10^{-2}~eV^2$,
which is not in the range accessible in these experiments.
However, this statement is
strictly true only to the extent that the leptonic mixing matrix is
assumed to be unitary.  Deviation from unitarity can produce oscillation
signals, although there should be no spatial dependence.  This situation
can arise if the ordinary neutrinos mix with a neutral heavy lepton with
mass on the order of the electroweak scale.  Let us
consider this in more detail.

For definiteness, let us assume that the three light neutrinos have an
admixture of a heavy fourth generation neutrino with mass greater than
$M_Z$.  Writing
\begin{equation}
\nu_e = \sum_{i=1}^4 U_{ei}\nu_i
\end{equation}
and similarly for the weak eigenstates $\nu_\mu, \nu_\tau$, one sees
that muon neutrinos produced in $\pi$ decays will be the weak eigenstates with
the heavy $\nu_4$ component removed.  The $\nu_e$ and $\nu_\mu$
states are no longer orthogonal, leading to an apparent oscillation
probability given by \cite{pakvasa}
\begin{eqnarray}
P_{\nu_e-\nu_\mu} &=& \left|(U_{e 1}^*U_{\mu 1}+U_{e 2}^*U_{\mu 2}+
U_{e 3}^* U_{\mu 3})\right|^2 \nonumber \\
&=& \left|-(U_{e4}^*U_{\mu4})\right|^2~.
\end{eqnarray}
Here we have set the $\Delta m^2$s to zero since they are all
$\mathrel{\raise.3ex\hbox{$<$\kern-.75em\lower1ex\hbox{$\sim$}}} 10^{-2}~eV^2$
and thus negligible for accelerator neutrino experiments.
If the mixing parameters $U_{e 4},U_{\mu 4} \sim ({1 \over 4} ~{\rm to}~
{1 \over 5})$, then the oscillation probability at the LSND experiment will be
in the range of $(2-3)\times 10^{-3}$ with no position dependence.  The data
presented in Ref. \cite{LSND}
is consistent with such an interpretation, as can be
inferred from the large $\Delta m^2$ region of their Fig. 3.  Various
other constraints on
such leakage to a heavy fourth generation
neutrino has been studied in \cite{babuma}, where it is shown that the
best existing limit
is from neutrino oscillation experiments.  $\nu_\mu-\nu_\tau$
oscillation probability of $(2-3) \times 10^{-3}$ can also be achieved by
a similar mechanism, but this may be more difficult to measure.

If the heavy neutrino is a standard model singlet, there are stringent
limits from $Z$--boson decays which make the non--unitary oscillations
unobservable at accelerators.  To see this, consider the mixing of
ordinary neutrinos with a heavy isospin singlet
neutral lepton $N$ of mass $m$.  In the
presence of such mixings, the invisible decay width of $Z$ will be
modified so that the number of effective neutrino species coupling to
$Z$ is given by
\begin{equation}
N_\nu = 3 + (1-|\alpha|^2)\left[-(1+|\alpha|^2)+2F|\alpha|^2+F'
(1-|\alpha|^2)\right]
\end{equation}
where $F$ and $F'$ are phase space factors given by
\begin{eqnarray}
F &=& 1 -{3 \over 2} x^2 + {1 \over 2} x^3 \nonumber \\
F' &=& \sqrt{1-4x^2}(1-x^2)
\end{eqnarray}
with $x\equiv m/m_Z$.  It is to be understood here that for $m \le m_Z/2
$, all terms in eq. (3) will contribute to $N_\nu$, while for
$m \ge m_Z$, the $F$ and $F'$ terms do not contribute.  If
$m_Z/2 \le m \le m_Z$, the term proportional to $F'$ should be set to
zero, while $F$ is non--zero.
The mixing parameter $|\alpha|$ is equal to the (4,4) element of
the unitary matrix $U_\nu$ which diagonalizes the neutrino mass matrix.  It
is related to the leptonic mixing matrix $U$ appearing in the
charged current via
\begin{equation}
|\alpha|^2 = 1-|U_{e4}|^2-|U_{\mu 4}|^2-|U_{\tau 4}|^2~.
\end{equation}

Although $U$, now a $3 \times 4$ matrix is not unitary,
the $\nu_e-\nu_\mu$ transition probability
is still given by eq. (2), since each row of $U$ is normalized to 1.  It
becomes apparent from eq. (2) and (5)
that $P_{\nu_e-\nu_\mu} \le (1-|\alpha|^2)^2/4$.
If the mass of $N$ is greater than $m_Z$, then $F=F'=0$, so
that $P_{\nu_e-\nu_\mu} \simeq [(3-N_\nu)/(1+|\alpha|^2)]^2/4
\equiv [\Delta N_\nu/(1+|\alpha|^2)]^2/4$.  $\Delta N_\nu$ can be as
large as 0.042 at 1 sigma \cite{PDT},
which when combined with $|\alpha|^2 \simeq 1$
yields $P_{\nu_e-\nu_\mu} \le 1.1 \times 10^{-4}$, too small to be
observable.  (Even with $\Delta N_\nu = 0.067$, the 2 sigma value,
$P_{\nu_e-\nu_\mu} \le 2.8 \times 10^{-4}$.)  If the mass of $N$ is
less than $m_Z/2$, then both the $F$ and $F'$ terms in eq. (3)
will be relevant, making $N_\nu$ more consistent with 3.  (As
$F,F' \rightarrow 1$, $N_\nu \rightarrow 3$.)  However, such a scenario
is ruled out by direct search limits for a heavy neutral lepton decaying
into the usual charged leptons $(e, \mu)$, as should happen here.  (The
limit on the mass of such a neutral lepton is $m \ge 46~GeV$).)
If the mass of $N$ obeys $m_Z/2 \le m \le m_Z$, then
$F \neq 0$, but $F'=0$ in eq. (3).
In this case, noting that $F$ is at most 11/16 (for $x=1/2)$, we
see that $P_{\nu_e-\nu_\mu}$ can be as large as $(4 \Delta N_\nu/5)^2
\simeq 10^{-3}$, which is close to present experimental sensitivity.
However, since the branching ratio  $Br[Z \rightarrow (N \overline{\nu} +
\overline{N} \nu)] \sim 2 \times 10^{-4}$, few hundred such events
should have been observed at LEP with
the $N$ decaying subsequently into $l^+ l^- \nu$.  This possibility is
also excluded based on the non--observation of such events.

We conclude that
the case of isosinglet neutrino does not lead to any observable
deviation from unitarity in the accelerator neutrino oscillation
experiments.  Note that there is no such constraint on a sequential fourth
generation neutrino (or neutrinos in vector--like families).

\noindent\underline{\it Scenario (ii): No degenerate neutrinos:}  If
there is no degeneracy in the neutrino masses and we assume some kind of a mass
hierarchy, accounting for
the solar, atmospheric and hot dark matter neutrinos would require the
introduction of a light (sterile) neutrino $\nu_s$.  There is then just one
consistent scheme in this case.  As far as we are aware, this possibility has
not been discussed in the literature, so we shall elaborate on it.


By assumption, in this scenario the known neutrinos are non--degenerate, and
indeed have hierarchical masses.  The
solar neutrino puzzle is resolved via $(\nu_e-\nu_s)$ MSW oscillations
such that $m_{\nu_s}^2-m_{\nu_e}^2 \sim 10^{-5}~eV^2$.  The atmospheric
neutrino anomaly is explained via $(\nu_\mu -\nu_e)$ oscillations with
$m_{\nu_\mu} \sim 10^{-1}~eV$.  The tau neutrino with a few eV mass
constitutes the `hot' component of dark matter.

As far as
$(\overline{\nu}_\mu-\overline{\nu}_e)$
oscillations at accelerators are concerned, the ``direct'' transition is not
possible  because $\Delta m^2 \sim 10^{-2}~eV^2$.  However, it has
recently been pointed out \cite{BPW}
that ``indirect'' $(\overline{\nu}_\mu-\overline{\nu}_e)$ transition
via a virtual $\overline{\nu}_\tau$ in the few eV mass range can still occur in
such a scheme.  For this to be in the experimental reach,
both $(\nu_e-\nu_\tau)$ oscillations (at reactors)
and $(\nu_\mu-\nu_\tau)$ oscillations (at accelerators) should be in the
observable range.  As shown in \cite{BPW}, the present limits from
Bugey reactor \cite{Bugey} on $\nu_e-\nu_\tau$ oscillations,
and the CHARM-II, Fermilab E531 and CDHS
limits \cite{E531} on $(\nu_\mu-\nu_\tau)$
oscillations translates into an observable ``indirect'' $(\nu_e-\nu_\mu)$
oscillations for a $\nu_\tau$ mass in the few eV range, which can
account for the LSND results.

Let us point out that in this second scenario, the allowed MSW parameter
region will be somewhat shifted because of the rapid $(\nu_e-\nu_\mu)$
oscillations from the sun to earth, leading to a further depletion of
$\nu_e$'s by a factor of
${1 \over 2} {\rm sin}^22\theta_{e\mu} \sim ({1 \over 4}~{\rm to}~
{1 \over 2})$.  This
prediction will be tested in the forthcoming solar neutrino experiments.

This scheme may be on a somewhat better theoretical footing than the first one
because no mass degeneracy is assumed. A light `sterile'
neutrino state, following \cite{bm},  may arise from the hidden sector of
some fundamental theory.   If the hidden sector contains $SU(3) \times SU(2)
\times U(1)$, one can define an  unbroken `parity' at the Lagrangian level.
The
$\nu_R$'s needed for the  see--saw mechanism, having no gauge quantum numbers
can freely mix from the  observable and hidden sectors.  This would lead to
mixing between the light neutrinos and their mirror partners (which are also
light).  Gravitationally induced interactions could also mix the neutrinos from
the two sectors.

\noindent \underline{\it Scenario (iii): Nearly degenerate $(\nu_\mu-
\nu_\tau)$ pair:}  This scheme also requires a light sterile neutrino
$\nu_s$.  In addition, an approximate lepton number symmetry such as
$L_e+L_\mu-L_\tau$ is necessary to make $\nu_\mu$ and $\nu_\tau$ nearly
degenerate in mass \cite{caldmohap,bm,maroy}.
When this symmetry breaks by a small amount, maximal
$(\nu_\mu-\nu_\tau)$ mixing occurs, facilitating a resolution of the
atmospheric neutrino problem.  The solar neutrino puzzle is explained via
$(\nu_e-\nu_s)$ oscillations.  The common mass of $\nu_\mu$ and $\nu_\tau$ is
assumed to be $\sim$ few eV, so that they constitute a `two--flavor'
hot dark matter.  For accelerator neutrinos, there is direct $(\nu_e-
\nu_\mu)$ oscillations at an observable level.  In particular, the
LSND data can be accommodated.  However, $(\nu_\mu-\nu_\tau)$ oscillations will
be beyond the reach of CHORUS/NOMAD and E803.

Aside from the two--flavor hot dark matter component, this scenario has
other special cases relevant for cosmology.  The solar neutrino data
requires the mass splitting $\Delta m^2$ between $\nu_e$ and
$\nu_s$ to be $\sim 10^{-5}~eV^2$, but the masses themselves may be in the
cosmologically interesting range of a few eV.  Since this pair
behaves like Dirac neutrinos, there is no conflict with
$\beta\beta_{0\nu}$ constraints.  For cosmology however, the
hot dark matter may be a linear combination of $\nu_e$ and $(\nu_\mu+
\nu_\tau)$.  If the total mass of $(\nu_\mu,\ \nu_\tau)$ pair exceeds that of
$\nu_e$, we call it a (2+1) C$+$HDM scheme, and if the $\nu_e$ mass dominates
over $(\nu_\mu,\ \nu_\tau)$, it is a (1+2) scheme.  Of course, one could
also recover the one flavor case by making the mass of
$(\nu_\mu-\nu_\tau)$ pair much smaller than 1 eV.  (This case will
correspond to an inverted mass hierarchy \cite{silk}.)

\section{Cosmological Implications}

   In studies of large scale structure, it has become clear that a critical
density Cold Dark Matter (CDM) universe with density perturbations that have
roughly a Harrison-Zeldovich spectrum cannot simultaneous fit observations of
structure on large and small scales.  Normalized on large scales to fit the
COBE observations, CDM produces too many clusters of galaxies and
galactic pairwise velocities far in excess of observations \cite{cdmnbody}.
This basic problem is neatly resolved in a cold plus hot
dark matter scenario because the growth of the density perturbations on small
scales is damped by the presence of the hot dark matter component
\cite{ss84,sss89,nbody}.  Since the neutrino oscillation experiments
may be telling us that the neutrino component is spread among several
flavors we would like to understand, in a quantitative way, how this changes
the usual (one flavor) cold plus hot dark matter model.  We will begin with a
general  description of the effects of a hot component on the growth of density
fluctuations.  First we will give some analytic formulae
for estimating the size of the effects, then we will show some results of
more detailed calculations.  Finally we conclude with some remarks about the
uncertainties in cosmological parameters which currently prevent one to
use cosmological data to pin down the values of the neutrino masses.

\subsection{ Free--Streaming or Jeans Masses}

 Neutrinos retain large
velocities from the time before primordial nucleosynthesis when they were in
thermal equilibrium.  This means that if one initially sets up density
perturbations, the neutrinos will rearrange themselves to a different pattern
at some later time, erasing neutrino density fluctuations on length scales over
which the neutrinos can have traveled.  This ``free--streaming" or ``Jeans"
length scale $\lambda_J$ is given at time $t$ by
\begin{equation}
\label{jeansl}
\lambda_J(t) = a(t) \tau (t) v(t),
\end{equation}
where $a(t)$ is the scale factor which describes the expansion of the universe,
$\tau(t)$ is the conformal time ($d\tau=dt/a(t)$), so $a(t) \tau (t)$ is the
physical horizon size, and $v(t)$ is the average neutrino velocity.  An
analytic fit for $v$ is given by \cite{bs83}
\begin{equation}
\label{neuvel}
v(t) = \left[ 1 + \left( {a\over a_{nr}}\right)^2 \right]^{-1/2},
\end{equation}
where $a_{nr}$ is the scale factor when the neutrinos become non-relativistic,
and we use units where the velocity of light is unity.  We have
\begin{equation}
\label{anr}
a_{nr} = 2.25 T_\gamma^0/m_\nu,
\end{equation}
where $T_\gamma^0 = 2.35\times 10^{-4}$ eV is the present cosmic background
photon temperature.  The value of the Jeans' length at the current time $t_0$
is
\begin{equation}
\label{lj0}
\lambda_J(t_0) = 3 t_0 a_{nr} = 3.2\left( {1\ {\rm eV}\over m_\nu}\right)
\ h^{-1}\ {\rm Mpc}.
\end{equation}
$h$ is the parametrization of our ignorance of the true value of the Hubble
constant $H_0 \equiv 100\ h$ km s$^{-1}$ Mpc$^{-1}$.  To relate this length to
the masses of known astrophysical structures we convert to the total
mass contained in the volume $4 \pi (\lambda_J/2)^3/3$, (the Jeans'
mass).  Currently, the Jeans mass $M_{J_\nu}(t_0)$ is
\begin{equation}
\label{mj0}
M_{J_\nu}(t_0) = M_{H_0} (a_{nr})^3 = 4.6\times 10^{12} \left( {1\ {\rm
eV}\over
m_\nu}\right)^3\ h^{-1}\ M_\odot,
\end{equation}
where $M_{H_0}= 3.14\times 10^{22}\ h^{-1}\ M_\odot$ is the mass contained in
the current Hubble volume.  Interestingly enough, the mass in (10) is
comparable
to  a galactic mass for neutrinos in the range $m_\nu \simeq 1-4$ eV.  Note
that
in the formulae presented here we assume a critical density of matter, and
only a fraction is in the form of hot dark matter-- the remainder is comprised
of cold dark matter and baryons.

  During the matter dominated epoch, $\rho\propto a^{-3}$, $v\propto a^{-1}$
(while the neutrinos are non-relativistic) and
$a \tau \propto a^{3/2}$, so the Jeans' mass $M_{J\nu} \propto a^{-3/2}$ and
decreases with time.  When the neutrinos are
relativistic, $v\sim 1$, and the Jeans' mass increases with time.  The
maximum in the Jeans mass occurs when the neutrinos are just becoming
non-relativistic, {\it i.e.}, when $a(t) = a_{nr}$:
\begin{equation}
\label{mjmax}
M_{J\nu}({\rm max}) \simeq M_{H_0} (a_{nr})^{3/2} = 4\times 10^{17}
\left( {1\ {\rm eV}\over m_\nu}\right)^{3/2}\ h^{-1}\ M_\odot.
\end{equation}
This relation is approximate, because the time when neutrinos become
relativistic is quite close to the time when matter begins to dominate over
radiation in the universe, so the horizon size does not scale exactly as
$a^{-3/2}$.  The exact formula must be determined numerically for each
value of the neutrino mass. The  formula in (11) is accurate to within an order
of magnitude, which is good enough for estimating the size of the effects under
study here.  In the older models of structure formation in which the dark
matter is totally composed of neutrinos of mass $\sim$ 30 eV, this maximum
Jeans mass is somewhat larger than a typical mass of a cluster of galaxies.

In studies of single massive flavor C$+$HDM models, a more typical value of
the neutrino mass $\sim 6$ eV (for $h=0.5$) was preferred, yielding a maximum
Jeans' mass of
$M_{J_\nu}(max) = 3\times 10^{16}\ h^{-1}\ M_\odot$, which is slightly smaller
than the so--called ``great attractor" \cite{samurai}, the largest
concentration of mass
which is pulling galaxies gravitationally on scales of order $40\ h^{-1}$ Mpc.
If we assume, as in scenario i), that the same $\Omega_\nu$ is now composed of
3 nearly degenerate neutrinos of mass $\sim 2$ eV,
$M_{J_\nu}(max) = 1 \times 10^{17}\ h^{-1}\ M_\odot$, which is somewhat
larger than the ``great attractor" mass.

    We will proceed to examine how these two mass scales ($M_{J\nu}(t_0)$ and
$M_{J\nu}(max)$) are incorporated in predictions of structure formation.

\subsection{Growth of Density Perturbations}

   In the early universe, when matter begins to dominate the energy density of
the universe, the CDM and C$+$HDM mass fluctuation spectra are identical.  The
only feature imprinted on the initial spectrum at this time is a decrease of
amplitude on length scales smaller than the horizon at matter domination.
These scales experience a smaller amount of growth during the radiation
dominated era due to the effects of radiation pressure.  After matter
domination the fluctuation amplitudes on these scales can grow, driven
by their own self--gravitation.  If a volume of space contains a mass
$\tilde{M}$ which is slightly different than the average $M$ for that size, and
the dark matter is strictly CDM, the fluctuation amplitude
will grow proportional to the scale factor $a$
\begin{equation}
\label{cdmgrowth}
{\delta M \over M}(M)\equiv {\tilde{M} - M \over M} \propto a
\end{equation}

Growth in the C$+$HDM model is more interesting.
When a mass fluctuation contains a mass $M<M_{J_\nu}(t)$ at time $t$, the
neutrinos will stream out of the fluctuation and change the local gravitational
potential.  This smaller potential also reduces the growth rate of the mass
fluctuations to \cite{ss84,rs89}
\begin{equation}
\label{chdmgrowth}
{\delta M \over M}(M) \propto a^p;\ p= {1\over 4}\left( -1 +
\sqrt{25 - 24 \Omega_\nu}\right)
\end{equation}

   The fact that the Jeans mass decreases with time means that if $M<M_{J_\nu}$
at some early time, sometime in the future $M>M_{J_\nu}$ will be true.  Once
$M>M_{J_\nu}$, the growth rate returns to the CDM growth rate, as the neutrinos
will then effectively be ``cold" on this mass scale.  This leads to scale
dependent changes in the mass fluctuation spectrum.  The equations for the
Jeans
mass (\ref{mjmax}) and (\ref{mj0}), and the growth rate
(\ref{chdmgrowth}) also illustrate the effect of having
multiple flavors of massive neutrinos with comparable masses.  The Jeans mass
depends on the individual neutrino mass, while the growth rate depends only on
the {\sl sum} of the neutrino masses, as
\begin{equation}
\label{omeganu}
\Omega_\nu =  h^{-2} \sum_{i= e,\mu,\tau} {m_{\nu_i} \over 94 \ {\rm eV}}.
\end{equation}
So increasing the number of degenerate flavors means that the Jeans mass will
be increased for a given value of $\Omega_\nu$.

   The growth of fluctuations on mass scale $M$ in the intermediate
region $M_{J_\nu}(max)<M<M_{J_\nu}(now)$ will be ``damped" relative to
their value in a universe dominated only by CDM.  The amount
of damping between the time when the neutrinos become non--relativistic and
when the Jeans mass becomes smaller than scale $M$ is given by
\begin{equation}
\label{damping}
{\rm damping\ factor} = \left[{M_{J_\nu}(max)\over M }\right]^{(2p
-2)/3}
\end{equation}
The maximum amount of damping occurs for mass scales $M\ll M_{J\nu}(t_0)$
which is a constant given by
\begin{equation}
\label{dampmax}
{\rm max\ damping\ factor} = \left[{M_{J_\nu}(max)\over
M_{J_\nu}(t_0) }\right]^{(2p-2)/3} = (a_{nr})^{1-p}
\end{equation}
The difference, then, between having the mass equally
spread among ${\cal N}_\nu$ flavors as opposed to 1 flavor results in an
additional small scale damping of ${\cal N}_\nu^{1-p}$, which is between
10\% - 20\% for $0.20< \Omega_\nu<0.30$.

   The combination of hot and cold components of dark matter thus has the
following effects, when compared to the a model with only a cold component.
First of all, the growth of density perturbations with
masses in the intermediate range $M_J(now)<M<M_J(max)$ is reduced by the factor
given in eq. (\ref{damping}) which depends mainly on $\Omega_\nu$.  Below
$M_J(now)$ the mass fluctuation spectrum follows the scale dependence of a CDM
model but with an amplitude reduced by the factor in eq. (\ref{dampmax}).
Splitting the neutrino density among ${\cal N}_\nu$ flavors of approximately
degenerate neutrinos produces an additional damping of ${\cal N}_\nu^{p-1}$
below a
somewhat higher value of $M_{J\nu}(max)$.  This additional damping may be
important for reconciling observations of clusters of galaxies with a COBE
normalized spectrum of density fluctuations.

\subsection{Detailed Comparisons to Structure Formation}

   In order to do a more careful comparison of these models to observed
large scale
structure, a more accurate evaluation of the effects expressed above is
required.  We achieve this by
integrating the linearized general relativistic evolution equations for the
photons, neutrinos (massive and effectively massless flavors), baryons and CDM
particles.  The procedure we use is described in
detail in ref. \cite{rsdelaix}.  The models have a baryon mass fraction
$\Omega_{baryon} = 0.0125/h^2$, which is reasonably consistent with
nucleosynthesis \cite{bbn}.  The CDM fraction is then given by $\Omega_{CDM} =
1-\Omega_\nu-\Omega_{baryon}$ (since here we are only considering critical
density universes).  In figures 1 and 2, we use a
Harrison--Zeldovich (scale free) spectrum normalized to COBE \cite{bennett94}
with $h=0.5$. After smoothing the density field with
a low pass Gaussian filter, we arrive at the curves shown in figures 1 and 2.
 The masses on the x-axis are $(2\pi)^{3/2}
r_f^3 \rho$, as appropriate for a Gaussian filter radius of $r_f$.
In the three panels a), b), and c), we show curves for $\Omega_\nu = 0.20$,
$0.25$, and $0.30$ respectively.  Within each panel we show curves for 1, 2,
and 3, degenerate neutrino flavors.  In panel a), we also display the analogous
CDM model curve for comparison.

The models  under consideration are now known to fit a
large amount of data on large scale structure
\cite{vdns,others,pogosyan93,ss94},
but here we will concentrate on
the two toughest constraints for C$+$HDM models to satisfy.   The first
constraint is on the amplitude of mass fluctuations which form clusters.
It has been argued \cite{white93} that the number density of rich
clusters requires that  on a scale
$\sim 10^{15}M_\odot$, $\delta M/M = 0.58\pm 0.1$ (2 times their ``error
bar").
 As stated in ref. \cite{white93} the errors are hard to estimate because of
possible systematic uncertainties in the masses assigned to clusters.
Indeed, the mass scales for two different determinations of the cluster mass
function in refs \cite{bahcall92} and \cite{biviano93} differ by almost a
factor of 2.  Ref. \cite{white93} argues that the cluster masses could
be overestimated, while measurements of the gravitational lensing by the
clusters suggest \cite{miralda93} that the usual mass estimates (used to
determine ($\delta M/M $)) may be low by as much as a factor of 3.
In order to allow for this uncertainty, we also give an estimated horizontal
error bar which is a factor of two (in cluster mass) from the value
$1\times 10^{15}\ h^{-1}\ M_\odot$, to reflect the size of the possible
systematic error.

   At the far left end of the figures we have also drawn a line which
corresponds to the lower limit of the mass fluctuation amplitude implied by the
measurements of the abundance of ``damped Lyman alpha systems" as indicated by
the recent survey of ref. \cite{storrie95}; see also ref. \cite{lymana}.
While early work  \cite{lymanth1}
indicated that C$+$HDM models (with $\Omega_\nu=0.3$) predictions were far
below
the observations, it now seems as though the assumptions used may have
been too restrictive.  There is little information about the
exact nature of
these damped Lyman alpha systems, other than their neutral hydrogen column
densities and spectral line widths.  A more general analysis \cite{klypin95}
shows how to properly estimate number abundances of objects on these scales
\cite{badlya}.  We have used this latter procedure to estimate the amplitude
of the constraint line in figures 1 and 2.  It is still uncertain what
mass is to be identified with these damped Lyman alpha
systems.  It seems reasonable that we should associate dark matter halo masses
of between $\sim 10^{10}-10^{11}\ h^{-1}\ M_\odot$ and we have placed our
constraint line there in our figures.  (The curves should pass above the
constraint line somewhere in the range.)  The mixed dark matter models
are more compatible with the lower end of this mass range.  However, with
$\Omega_\nu = 0.30$ and $h=0.5$ one has to take very small halo masses
even to make this model work marginally.  From figure 1 then it is apparent
that $\Omega_\nu <0.30$ if $h=0.5$ with a pure Harrison--Zeldovich
spectrum.

  Figure 1 also points out an interesting aspect of spreading the neutrino mass
density among a number of massive neutrino flavors ${\cal N}_\nu$.  With
${\cal N}_\nu=1$, the amplitude of the
density fluctuations on cluster scales seems to be slightly larger than what is
required; increasing ${\cal N}_\nu$ improves the fit.  This result depends only
weakly on $\Omega_\nu$ on this mass scale. [Eq.(\ref{damping}) predicts less
than 10\% difference in the cluster scale $\delta M/M$ amplitude over the range
$\Omega_\nu \sim 0.2-0.3$.]  However, because of the
$\Omega_\nu$ dependence in the ${\cal N}_\nu^{p-1}$ damping factor (via $p$),
${\cal N}_\nu=2$ works best for $\Omega_\nu=0.30$ and ${\cal N}_\nu=3$ works
best for $\Omega_\nu=0.20$.  On the other hand increasing the number of
flavors makes the disagreement with Lyman alpha systems datum worse for
$\Omega_\nu=0.30$, so ${\cal N}_\nu>1$ seems to work best for
models with $\Omega_\nu<0.30$.

We should point out here that the conclusions
we draw from figure 1 hold only for the spectrum, Hubble constant and baryon
fraction used in the models, and altering any of these parameters changes the
conclusions.  We will explore this further later.  The purpose of figure 1 is
mainly to illustrate the effects of modifying the hot dark matter composition.

\subsection{Additional Models Which Can Accommodate the LSND Results}

    In the above we have considered the consequences of the
three neutrino scenarios in which we have ${\cal N}_\nu$ flavors with
roughly equal mass.   However, if we take the LSND results as a serious
constraint, then scenario (iii) admits the possibility that all three
known neutrino flavors can have masses in the eV
range, although the electron neutrino mass can still be significantly different
from the $\mu$ and $\tau$ neutrino masses (which form a degenerate
pair).  Since only the square of the mass difference is
measured,
we have two cases to consider: a) $m_{\nu_\mu}>m_{\nu_e}$, (``2+1") and
b) $m_{\nu_e}>m_{\nu_\mu}$ (``1+2").

In a ``$2+1$" model, where\ $\nu_\mu$ and
$\nu_\tau$ are each heavier than $\nu_e$, the lightest neutrino ($\nu_e$) will
have a very large maximum Jeans' mass, ($M_{J_\nu}^{l}(max)$), compared to
the other two heavier neutrino Jeans masses ($M_{J_\nu}^{h}(max)$).  In
the ``$1+2$" model, where$\nu_\mu$ and $\nu_\tau$ are lighter than
$\nu_e$, the lightest neutrinos ($\nu_\mu$ and $\nu_\tau$) will
have very large $M_{J_\nu}^{l}(max)$, compared to the remaining heavier
$\nu_e$ neutrino $M_{J_\nu}^{h}(max)$.
For $M_{J_\nu}^{l}(max)>M>M_{J_\nu}^{h}(max)$, there will be an
amount of damping which would be there if only the lightest neutrino comprised
$\Omega_\nu$.  For $M_{J_\nu}^{h}(max)>M$ the damping will be intermediate
between the one and three flavor massive neutrino case.

To illustrate these two cases, we will
use values of $\Delta m^2=8$ and 20 eV$^2$, for the ``$2+1$" and ``$1+2$"
models respectively in the context of an $\Omega_\nu=0.25$, $h=0.5$, $n=1$
cosmology.   In figure 2 we plot the curves for the $2+1$ model
($m_{\nu_e}=0.22$ eV and $m_{\nu_\tau},m_{\nu_\mu} = 2.84$ eV), and the
$1+2$ model ($m_{\nu_e}=4.524$ eV and $m_{\nu_\tau},\ m_{\nu_\mu} = 0.685$ eV.
For comparison we also plot the curves for ${\cal N}_\nu=1$, 2, and 3 flavors
for the same $\Omega_\nu\ (=0.25)$. The curve for $2+1$ ($1+2$) lies in
between the curves for the 2 and 3 (1 and 2) degenerate flavors.

   We can estimate the difference between having 3 flavors with equal and
unequal masses as follows.  In the ``1+2" and ``2+1" models the lightest
flavor mass $m_{\nu l}$ is lighter than the three equal flavor mass
$m_{\nu 3}$.  The heavier flavor mass $m_{\nu h}$ is greater than
$m_{\nu 3}$.  Consider a mass scale $M<M_{J_\nu}^{h}(max)$.  After some time,
when $M_{J_\nu}^{h}(t)= M$, the amount of damping is the same in both the equal
and unequal mass cases.  After that time that scale will experience an
additional damping of $[M_{J_\nu}^l(max)/M_{J_\nu}^{h}(max)]^{(2p_l
-2)/3}$ (eq. \ref{damping}) in the unequal mass case, and
$[M_{J_\nu}^{(3\ flavors)}(max)/M_{J_\nu}^{h}(max)]^{(2p-2)/3}$ in the 3 flavor
case. Here $p_l$ is $p$ in equation (\ref{chdmgrowth}) calculated assuming
$\Omega_\nu$ is only made from the lightest neutrino(s).  The ratio of the
damping in the 1+2 (and 2+1) cases to the 3 equal mass case is (for scales
$<M_{J_\nu}^{h}(max)$)
\begin{equation}
\label{dratio}
{\rm damping\ ratio} = { (m_{\nu l}/m_{\nu h})^{(1-p_l)} \over
(m_{\nu 3}/m_{\nu h})^{(1-p)} },
\end{equation}
The ratio in (\ref{dratio}) is
always larger than 1, indicating that the damping is greatest in the 3 equal
flavor case.  However, if the ratio of the masses of heavy to light flavors
is $\mathrel{\raise.3ex\hbox{$<$\kern-.75em\lower1ex\hbox{$\sim$}}} 4$ the
difference between damping in the various cases differs from the 3 flavor case
by $<2$ \%.  This is the reason for choosing the particular $\Delta m^2$ for
plotting.  Had we plotted, for example, the 2+1 and 1+2 models using
$\Delta m^2 =6$ eV$^2$, (the same value used in  ref. \cite{primack95}, but
for $\Omega_\nu=0.20$) both of the curves would be nearly indistinguishable
from
the  3 degenerate neutrino case for masses $<M_{J_\nu}^h(max)$.
When the light and heavy masses differ by an order of magnitude or more, as in
figure 2, the $2+1$ model goes to the 2 degenerate flavor model and the $1+2$
goes to the single flavor model. We note that because the lightest neutrino
weighs less than the 3 degenerate flavor neutrino, on the very largest scales
where $M^l_{J_\nu}(max)<M<M^{(3\ flavors)}_{J_\nu}(max)$, there will be some
damping which is absent in the 3 degenerate flavor case.

We see that these additional models offer now a continuum of $\Delta M/M$
values between the 1 and 3 degenerate mass neutrino cases. The sole motivation
for considering a case where the neutrinos have masses {\em and} mass
differences of order a few eV in scenario iii) is the possibility of detecting
oscillations with $\Delta m^2$ in the few eV$^2$ range in the LSND (and also
KARMEN) experiments. In the framework of GUT theories however, such a mass
spectrum will not be easy to understand.

\subsection{Sensitivity to Cosmological Parameters}

   In the previous sections we have plotted results for specific values of the
Hubble constant, initial density fluctuation spectrum, and $\Omega_{baryon}$.
We have not drawn strong conclusions about the
neutrino mass scenarios, because the results are quite sensitive to the
specific values used for these cosmological parameters.  We will discuss the
effects of these parameters on the cosmological structure formation and give
specific examples of models from each of the three mass scenarios.

\begin{enumerate}
\item{$H_0$:}  The value of the Hubble constant has been the subject of
a long and ongoing observational campaign (see \cite{hubble} for some recent
measurements).  The experiments
find that $0.4<h<1.0$.  This uncertainty is quite large when seen from the
point of view of the required neutrino masses.  Rewriting equation
(\ref{omeganu}) for degenerate neutrinos we find
\begin{equation}
\label{numass}
m_\nu = 4.7 {1\over {\cal N}_\nu} {\Omega_\nu \over 0.2} \left({h\over
0.5}\right)^2\
{\rm eV},
\end{equation}
which emphasizes that the neutrino masses scale as $h^2$.  In
figures 1 and 2 we have used $h=0.5$.  Some of the recent measurements
\cite{hubble} seem to imply a larger value of $h$.  In particular, if
$h=0.6$, then the masses used in the previous examples are increased by a
factor of $1.44$.  For example, ref. \cite{primack95} advocates using
${\cal N}_\nu = 2$ and $\Omega_\nu = 0.2$ with $\Delta m^2 = 6$ eV$^2$,
$h=0.5$.  If instead $h=0.6$ with the other parameters fixed, then the
oscillation signal at Los Alamos would be consistent with
$\Delta m^2 = 12$ eV$^2$.

   The scale which corresponds to the horizon size at matter domination also
depends on the Hubble constant.  If we use a larger value of $h$, then matter
domination occurs earlier, when the horizon size was smaller, which means that
the radiation dominated era is less effective at arresting the growth of small
scale fluctuations.  Such an effect exacerbates the small scale problems in
CDM models and favors using a larger value of $\Omega_\nu$
\cite{pogosyan95,dss}.

are attributed

\item{ Initial Mass Fluctuation Spectrum}

It has long been known that inflation predicts a power spectrum of density
fluctuations with a spectral index $n$ close to unity.  Since the amount of
deviation is strongly model dependent, many investigators are content to use
$n=1$.  In figures 1 and 2 we have  also used the value of
$n=1$ for the spectral index.  However, even small deviations from $n=1$ can
lead to significant changes in the conclusions.  To see this we give the
initial spectrum in terms of rms mass
fluctuations (${\Delta M\over M}$):
\begin{equation}
\label{masspect}
{\Delta M\over M} \propto M^{-(3+n)/6}
\end{equation}
  In the simplest models of inflation, particularly those based on GUTs (see,
{\it e.g.,} ref. \cite{ss94,dss}), $n\sim 0.94-0.98$, although
other values of $n$ are certainly possible
\cite{inflrev}.  From eq. (\ref{masspect}) we see that decreasing
(increasing) $n$ from unity decreases (increases) the small scale power.  Since
the models are normalized at large scales to COBE observations the mass
fluctuation curves in figures 1 and 2 will ``pivot" around a very large mass
scale $\sim 10^{21}\ h^{-1}\ M_\odot$.  The ``pivot" mass scale here is
somewhat smaller than the horizon mass, because the best fit COBE quadrupole
anisotropy scales as $e^{1-n}$ \cite{bennett94}.  (We have fit our spectra by
normalizing to the 7th multipole moment of the Sachs-Wolfe anisotropy, as
recommended in ref. \cite{bennett94})

\item{$\Omega_{baryon}.$}
   In the past few years comparisons of the primordial light element abundances
inferred from observations with those calculated have led to strict
limits \cite{bbn} on the amount of baryonic material in the
universe, $0.04 <\Omega_{baryon} (0.5/h)^2 <0.06$.  We have used the central
value of this range for the models in figures 1 and 2.  Improvements in the
observations of deuterium and $^3$He and determinations of the neutron lifetime
have now led to a situation where all of the light element abundances are not
consistent in a universe where the three known neutrino flavors have the
standard  number densities (derived from thermal equilibrium) at
nucleosynthesis
\cite{steigman95}.  If, however, the $^4$He abundances derived from
observations have been systematically underestimated \cite{copi94} by about
5\%-10\% then big bang
nucleosynthesis would be made consistent with the three flavors of neutrinos,
provided that the baryon density is about 50\% larger than previously thought.
This would go in the right direction to explain why the baryon to dark
matter ratio in cluster cores is so much larger than $\Omega_{baryon}$
\cite{whitecoma94}.

If we allow for larger baryon fractions, this will also change the amplitude of
the mass fluctuations for a given $\Omega_\nu$ and $n$.  We can understand this
through the following.  Baryons (mostly protons) after nucleosynthesis are
electrically charged and so are strongly coupled to
the photon field.  Baryonic
density fluctuations cannot grow until the photon temperature decreases to
allow the stable formation of neutral hydrogen.  Because photon--baryon
decoupling does not happen until after matter domination, the baryons are
prevented from falling into the gravitational wells supplied by the cold dark
matter.  This leads to a damping in the growth of density fluctuations
(relative to a case with no baryons).  The amount of damping (on scales
smaller than the horizon size at photon baryon decoupling) in the final
amplitude of density fluctuations is constant, so the effect is quite similar
to changing the number of degenerate neutrino flavors.
\end{enumerate}

   These uncertainties mean that one cannot currently use the cosmological data
to determine which of our three scenarios is correct, or even what the value of
$\Omega_\nu$ is.  One must do a systematic study of the available parameter
space taking into account the full range of these parameters.  Studies of the
interplay between these cosmological parameters have been done in various
contexts \cite{pogosyan93,ss94,liddle93}, with more in progress
\cite{liddle95}.

   As a graphic illustration of this uncertainty, and also to demonstrate use
of the formulae presented in sections 3.1-3.3, we show four models in figure
3 that give nearly identical predictions of structure formation, but use
quite different solutions of the neutrino oscillation data.  The choices of
parameters are as follows.

\begin{enumerate}

\item  The solid curve.  [Scenario (iii)].  This model with
${\cal N}_\nu=2$, $\Omega_\nu=0.20$, $n=1$, and $h=0.5$ is
advocated as a good fit to cosmological observations by ref.
\cite{primack95} consistent with the Los Alamos LSND experiment.
We present this model for comparison.

\item The dot-dashed curve. [Also scenario (iii)].  We increase $h$ to
$0.6$ in this ${\cal N}_\nu=2$ model, but now there will be too
much small scale power.  To compensate for this, we increase $\Omega_\nu$ to
0.25, and decrease $n$ to 0.95, a value which actually is more in line with
standard inflationary predictions.  The $\bar{\nu}_e-\bar{\nu}_\mu$
experiments would then be expected to see a
signal corresponding to $\Delta m^2 = 18$ eV$^2$.
The baryon fraction has been scaled from the first model as $h^2$ consistent
with the nucleosynthesis constraints.

\item The short dashed curve. [Scenario (i)].  Again using $h=0.6$ and
$\Omega_\nu$, we now consider ${\cal N}_\nu=3$ degenerate flavors.
Because the extra flavor provides an additional damping factor over the two
flavor model of $(2/3)^{.16}= 0.94$, (eq. \ref{dampmax}), to get a curve
similar to the previous one we can increase $n$ to 0.98, which
increases the amplitude at the cluster mass by $(10^{21}/10^{15})^{0.0075} =
1.10$, (eq.
\ref{masspect}), so we get very nearly the same fit at the cluster scale.  The
value $n=0.98$ happens to be the value predicted in a particular SUSY GUT
inflation model \cite{dss}.

\item The long dashed curve. [Scenario (ii)].  Here we return to $h=0.5$,
$\Omega_\nu=0.2$, and $n=1$ as in the solid curve.  To compensate for the
increase in amplitude by going back to 1 flavor $(2/1)^{0.13} = 1.09$ in the
solid curve, we increase
$\Omega_{baryon}$ to 0.10.  This value of $\Omega_{baryon}$ is consistent with
3 relativistic neutrino flavors during nucleosynthesis if the $^4$He
abundances have been systematically underestimated.
\end{enumerate}

   We have now shown that allowing for more than one degenerate (in mass)
neutrino state, as indicated by solar and atmospheric oscillation experiments,
produces a ``degeneracy" in the predictions of structure formation for C$+$HDM
models, given the uncertainty in cosmological parameters.  In order to break
this latter ``degeneracy" we need a convincing detection of neutrino
oscillations in an
accelerator experiment, or an improvement in the determination of cosmological
parameters, or both.  In the meantime, we just note that there is a very rich
structure contained in C$+$HDM models of structure formation, which still
allows considerable flexibility in fitting astrophysical data.

\section{Conclusions}

 We have investigated the impact of neutrino oscillations, as indicated
by a number of solar and atmospheric neutrino experiments, on the
`cold plus hot' dark matter scenario of large scale structure formation.
We are led to three distinct scenarios for neutrino masses and mixings with
interesting predictions for the ongoing/planned experiments.  We note in
particular the expectations for $\bar{\nu}_\mu- \bar{\nu}_e$ and
$\nu_\mu-\nu_\tau$ oscillations being currently searched for.
  The cosmological implications of the three scenarios are explored
in some detail.   For some choices of
the cosmological parameters (particularly $h=0.5$, $n=1$, and
$\Omega_{baryon}=0.05$), the two and three neutrino flavors scenarios provide
a somewhat better fit to the
present data than the single (neutrino) flavor case.  However, as we show, this
need not hold for a different parameter choice.  It is too early to pick
out the best model but it is remarkable that taking account of the oscillation
experiments, the C$+$HDM models can still provide a good fit to the large scale
structure data within the context of the simplest inflation models.

\newpage
\section{ Figure Captions}

\noindent {\bf Fig. 1}.  We show the difference in the rms mass fluctuation
amplitude ($\Delta M/M$) between having the hot dark matter
(neutrino) component distributed among 1, 2, or 3 flavors of degenerate
neutrinos which correspond to our scenarios ii), iii) and i) respectively, for
fitting the neutrino oscillation data.  Here
$h=0.5$, $\Omega_{baryon}=0.05$, and $n=1$, with the spectrum normalized to
COBE.    The data points are: 1) the mass fluctuation
estimate based on cluster formation \cite{white93} and 2) the lower limit on
the mass fluctuations consistent with the formation of Lyman alpha cloud
abundances at high redshift \cite{storrie95}.  (The mass at which this lower
limit is to be imposed should be somewhere along the constraint line.)
Panels a), b), and c) correspond
to $\Omega_\nu=0.20,\ 0.25$ and 0.30, respectively.  It can be seen that
increasing the number of neutrino flavors marginally improves the fit to the
cluster data.  In panel a) we also show the curve for a pure cold dark matter
model.
\vspace{.6 cm}

\noindent {\bf Fig. 2}.  The rms mass fluctuation amplitude in a special case
of neutrino scenario iii, which has been made consistent with the LSND best
fit of
$\delta m^2 = 6$ eV$^2$ in a model with $\Omega_\nu = 0.25$, $h=0.5$, $n=1$.
For comparison we also show (from top to bottom) the same $\Omega_\nu$ with 1,
2, or 3 degenerate neutrino flavors.  The ``$2+1$" and ``$1+2$" model curves
are intermediate to the degenerate neutrino cases.  The data are the same as
in Figure 1.
\vspace{.6 cm}

\noindent {\bf Fig. 3}.  The rms mass fluctuation in some selected models with
varying cosmological parameters.  We plot curves for two models from scenario
iii with ${\cal N}_\nu=2$ for ($\Omega_\nu=0.20$, $h=0.5$, $n=1$) - as
advocated in  ref. \cite{primack95} - and for ($\Omega_\nu=0.25$, $h=0.6$,
$n=0.95$).  These two models would predict $\delta m^2 = 5.5$ and 18 eV$^2$
at the LSND and KARMEN experiments.  The third curve is for a model
with ${\cal N}_\nu=3$
($\Omega_\nu=0.25$, $h=0.6$, $n=0.98$).  Lastly we show a curve from scenario
ii) with ${\cal N}_\nu$ ($\Omega_\nu=0.20$, $h=0.5$, $n=1.00$).  Here we have
increased $\Omega_{baryon}$ to reduce the ampitude at cluster scales.  As can
be seen, it is difficult to
determine which neutrino mass scenario is correct based on current cosmological
data alone.  The data are the same as in Figure 1.


\begin{thebibliography}{abc}
\bibitem{also}{ Note that 3 degenerate flavors was also considered in refs.
\cite{sss89}, \cite{holtzman89} and the last article in ref. \cite{vdns}. }
\bibitem{badlya}{ Note that Ma and Bertschinger (last ref. in \cite{nbody})
do not entirely agree with this conclusion.  The disagreement relates to the
different algorithms for identifying dark matter halos in the simulations.}
\end{thebibliography}
\end{document}